\documentclass{article}
\catcode`\@=11 \@addtoreset{equation}{section}  
\renewcommand{\theequation}                     
         {\arabic{section}.\arabic{equation}}   

 \newcommand{\Lom}{\mathcal{L}}

\title{Geometry of the Divergences Problem in QFT}
\author{P. Leifer }
\date{Bar-Ilan University$^1$,  Ramat-Gan, Israel}
\begin{document}
\maketitle
\footnotetext[1]{On leave from Crimea State Engineering
and Pedagogical University, Simferopol, Crimea, Ukraine}
\begin{abstract}
The divergences problem in QFT should be overcame presumably due to
the unification of the fundamental interactions \cite{Dirac1}. We
evidently cannot to achieve this goal now. Together with this there
are divergences in problems where the high-energy processes simply
cannot be involved (say electron self-energy, Lamb's shift, etc.).
Last ones reflect the formal character of perturbation theory
applied to interacting secondly quantized amplitudes of the
pointwise particles. These difficulties are borrowed partly from the
classical theory. I would like to establish some general framework
for the future unified theory avoiding to use the method of
classical analogy.
\end{abstract}
\noindent PASC: 03.65.Ca, 03.65.Ta, 04.20.Cv
\section{Introduction}
The classical material points moving in space-time are main original
objects of the constructive quantum mechanics (QM) and quantum field
theory (QFT). This approach bring a lot of conceptual and technical
problems.  All progress of the theoretical physics demonstrates that
the pointwise in space-time interaction is approximate: the
mechanical interaction of solid bodies is the photon exchanges,
nuclear interaction provided by pions, and the pointwise Fermi
interaction realized by the $W^{\pm},Z^0$ exchange in electro-weak
theory. We may assume that all attempts to localize matter in
space-time, i.e. increasing of resolution by higher energy of
collision leads in fact to the delocalization by defreezing new
internal degrees of freedom. On the other hand, mathematically, this
locality leads to the singular functions (the most serious artifact
of the local QFT).

The modern nonlocal objects like strings and membranes arose due to
attempts to avoid these difficulties. Since physical status these
objects is not clear up to now and their logic leads far away from
ordinary physical paradigm \cite{W1}, it is worth while do find a
different solution of the divergences problematic. I show here
another method of introduction of nonlocal objects arising in
dynamical space-time which is built as specific section in the
tangent fibre bundle over the ``vacuum landscape''.

\section{Geometry of the Divergences Problem}
I start with simple Dirac's example of the divergences problem
\cite{Dirac1}. Dirac took the fermionic model Hamiltonian $
\hat{H}=\frac{1}{2}(a_{mn}\eta_m \eta_n - \bar{a}_{mn} \bar{\eta_m}
\bar{\eta_n})$. It is assumed that $\bar{\eta_n}|S> = 0$ for the all
$1\leq n \leq \infty$, where $|S>$ is a ``standard'' vector. The
matrix $a_{mn}$ is defined by Dirac as follows: $a_{mn} =
\delta_{m+1,n} - \delta_{m,n+1}$. It is easy to see $ \hat{H}^2|S> =
-\frac{1}{2}Tr(\bar{a}a) |S> = \infty |S>$. It should be noted that
the `deformation' of the standard state vector $|S>$ in this
artificial example has only `longitudinal' character, i.e. only
phase of the state vector $|S>$ is changed, not its direction.
Formally one can find Schr\"odinger solution with this Hamiltonian
$|\Psi(t)> = cos(\frac{t}{\hbar}Tr(\bar{a}a))|S>$, corresponding to
the infinitely fast phase oscillation. Definitely physicist has some
aversion to such behavior, but from the point of view of the
projective geometry (and postulate of the ordinal quantum mechanics)
`deformed' vector $|\Psi(t)>$ belongs to the same ray as $|S>$, i.e.
this is the same quantum state. This gives us the main idea to avoid
the divergences problem. Namely, the orthogonal projection along of
the vacuum  state is the subtraction the `longitudinal' component of
the variation velocity of the action state $|\Psi(t)> $.  Let put
$\eta_n$ to be creation operator and $|\xi_n> = \eta_n |S> =
\alpha_n |S > + |n)$ is deformed standard vector, so that $(n|S
>=0$, therefore, $<S |\xi_n> = <S | \eta_n |S > = \alpha_n <S |S >$
and, hence, $\alpha = \frac{<S | \eta_n |S >}{<S |S >}$. Now we can
express the `transversal' part of the standard vector deformation
\begin{eqnarray}{\label{1}}
|n) = |\xi_n> - \frac{<S | \xi_n>}{<S |S>}|S >,
\end{eqnarray}
so that it is orthogonal to $<S |$. Let me calculate only the
`transversal' components of the
$|\Psi(t)>=\exp(\frac{i}{\hbar}\hat{H}t)|S>$ which I will define as
follows:
\begin{eqnarray}
|\Psi(t)) = |\Psi(t)> - \frac{<S | \Psi(t)>}{<S |S >}|S
>=\exp(\frac{i}{\hbar}\hat{H}t)|S>-\frac{|S> <S|}{<S |S >}\exp(\frac{i}{\hbar}\hat{H}t)|S>
\end{eqnarray}
Now I apply this definition to calculation of all orders of
$|\Psi(t)>$:
\begin{eqnarray}
|\Psi(t)_0) = |S> - \frac{<S | S>}{<S |S >}|S >=|S>-\frac{|S>
<S|}{<S |S
>}|S>=0,
\end{eqnarray}

\begin{eqnarray}
|\Psi(t)_1) = t(\hat{H}|S> - \frac{<S |\hat{H}| S>}{<S |S >}|S
>)=t(\hat{H}|S>-\frac{|S> <S|}{<S |S
>}\hat{H}|S>)=t\hat{H}|S>,
\end{eqnarray}

\begin{eqnarray}
|\Psi(t)_2) = \frac{t^2}{2!}(\hat{H}^2|S> - \frac{<S |\hat{H}^2|
S>}{<S |S
>}|S>)
=\frac{t^2}{2!}(\hat{H}^2|S>-\frac{|S> <S|}{<S|S>}\hat{H}^2|S>) \cr
= \frac{t^2}{2!}(\frac{1}{4}(a_{mn}a_{pq}\eta_m \eta_n \eta_p \eta_q
-\bar{a}_{mn} a_{pq} \bar{\eta}_m \bar{\eta}_n \eta_p \eta_q)|S> \cr
-\frac{|S><S|}{<S|S>}\frac{1}{4}(a_{mn}a_{pq}\eta_m \eta_n \eta_p
\eta_q -\bar{a}_{mn} a_{pq} \bar{\eta}_m \bar{\eta}_n \eta_p
\eta_q)|S>) \cr = \frac{t^2}{2!}(\frac{1}{4}(a_{mn}a_{pq}\eta_m
\eta_n \eta_p \eta_q -2 Tr(\bar{a}a))|S> \cr -\frac{|S>
<S|}{<S|S>}\frac{1}{4}(a_{mn}a_{pq}\eta_m \eta_n \eta_p \eta_q -2
Tr(\bar{a}a))|S>) \cr =\frac{t^2}{2!}\frac{1}{4}a_{mn}a_{pq}\eta_m
\eta_n \eta_p \eta_q |S>,
\end{eqnarray}
where terms, giving nil contribution were omitted. We can see that
divergent second order term was canceled out. Let now see what
happen in the third order. It is easy to see that
\begin{eqnarray}
\hat{H}^3|S>=\frac{1}{8}(a_{mn}\eta_m \eta_n -\bar{a}_{mn}
\bar{\eta}_m \bar{\eta}_n )(a_{pq}\eta_p \eta_q- \bar{a}_{pq}
\bar{\eta}_p \bar{\eta}_q)(a_{rs}\eta_r \eta_s -\bar{a}_{rs}
\bar{\eta}_r \bar{\eta}_s)|S> \cr =\frac{1}{8}(a_{mn}\eta_m \eta_n
a_{pq}\eta_p \eta_q a_{rs}\eta_r \eta_s - a_{mn}\eta_m \eta_n
a_{pq}\eta_p \eta_q \bar{a}_{rs} \bar{\eta}_r \bar{\eta}_s \cr
-a_{mn}\eta_m \eta_n \bar{a}_{pq} \bar{\eta}_p \bar{\eta}_q
a_{rs}\eta_r \eta_s+a_{mn}\eta_m \eta_n \bar{a}_{pq} \bar{\eta}_p
\bar{\eta}_q \bar{a}_{rs} \bar{\eta}_r \bar{\eta}_s \cr
-\bar{a}_{mn} \bar{\eta}_m \bar{\eta}_n a_{pq}\eta_p \eta_q
a_{rs}\eta_r \eta_s+ \bar{a}_{mn} \bar{\eta}_m \bar{\eta}_n
a_{pq}\eta_p \eta_q \bar{a}_{rs} \bar{\eta}_r \bar{\eta}_s \cr +
\bar{a}_{mn} \bar{\eta}_m \bar{\eta}_n \bar{a}_{pq} \bar{\eta}_p
\bar{\eta}_q a_{rs}\eta_r \eta_s - \bar{a}_{mn} \bar{\eta}_m
\bar{\eta}_n \bar{a}_{pq} \bar{\eta}_p \bar{\eta}_q \bar{a}_{rs}
\bar{\eta}_r \bar{\eta}_s)|S> \cr =\frac{1}{8}(a_{mn}\eta_m \eta_n
a_{pq}\eta_p \eta_q a_{rs}\eta_r \eta_s-2 Tr(\bar{a}a) a_{mn}\eta_m
\eta_n)|S>,
\end{eqnarray}
and, therefore,
\begin{eqnarray}
|\Psi(t)_3) = \frac{t^3}{3!}(\hat{H}^3|S> - \frac{<S |\hat{H}^3|
S>}{<S |S
>}|S>) \cr
=\frac{t^3}{3!}\frac{1}{8} (a_{mn}a_{pq}a_{rs}\eta_m \eta_n \eta_p
\eta_q \eta_p \eta_s-2 Tr(\bar{a}a) a_{mn}\eta_m \eta_n)|S>.
\end{eqnarray}
One may see that the divergences alive in the third order. Since the
indefinite trace $Tr(\bar{a}a) = - \infty$ is a coefficient before
the transversal to the vacuum two-fermionic term, the compensation
projective term does not help. Nevertheless, we can extract the
useful hint: the vacuum vector (the standard vector in Dirac's
example) should be smoothly changed, and, furthermore,  the
transversal component should be reduced during the ``smooth''
evolution. One may image some a smooth surface with a normal vector,
taking the place of the vacuum vector. Then the orthogonal
projection acting continuously is in fact the covariant
differentiation of the tangent Hamiltonian vector field
\cite{Thorp}. One has in fact the modification of the
creation-annihilation operators of quantum particles. Let me recall
that the main technical result of Dirac approach \cite{Dirac1} is
the calculations of the coefficients $Y_{in}$ and $Z_{in}$ modifying
the initial creation-annihilation operators.

Of course, the trivial example of Dirac has only pedagogical sense
and one should take some more realistic model. Dirac undertook
attempts to find out the better approximation to the vacuum state
vector of the QED \cite{Dirac2}. The main idea was to safe physical
meaning of the Schr\"odinger picture in the QED. The approximate
state vector found by Dirac comprises of `longitudinal' and
`transversal' components corresponding creations electron-positron
pears and the two electron-positron pears. Thereby `deformed' vacuum
vector has not only trivial modulus variation, but the variation of
a direction too. Both of these approximate vectors $|V_1>$ and
$|V_2>$ suffer on divergences and Dirac came to the conclusion, that
the Schr\"odinger vector in QED does not exist at all, because it
does not lie in any separable Hilbert space \cite{Dirac1}. In spite
of this extreme point of view, I will assume that a tangent vector
of state `creeps' along the projective Hilbert space $CP(N-1)$ from
one to another tangent Hilbert spaces $T_{\pi}CP(N-1) = C^{N-1}$ at
different generalized coherent states (GCS) serving for the
parametrization of quantum setup \cite{Le1}. Hereafter I will use
finite dimension case $SU(N)$, $CP(N-1)$, etc., but where it is
necessity, the limit $N \to \infty$ is keeping in mind. This
modification requires, of course, a deep reconstruction of the QFT
\cite{Le2,Le3,Le4,Le5,Le6}. Physically main content of this program
is to design all physical entities in the terms of the pure quantum
state space geometry. Many inequivalent representations realized now
in the tangent fibre bundle over $CP(\infty)$.
\section{Action projective state space}
Hereafter (beside the paragraph 4) I will use the indices as
follows: $0\leq a \leq N$, and $1\leq i,k,m,n,s \leq N-1$

One of the most serious modification concerns the scheme of the
``second quantization'' procedure.

{\it First.} In the second quantization method one has formally
given particles whose properties are defined by some commutation
relations between creation-annihilation operators. Note, that the
commutation relations are only the simplest consequence of the
curvature of the dynamical group manifold in the vicinity of the
group's unit (in algebra). Dynamical processes require, however,
finite group transformations and, hence, the global group structure.
The main technical idea is to use vector fields over group manifold
instead of indefinite Dirac's q-numbers. This scheme therefore
looking for the dynamical nature of the creation and annihilation
processes of quantum particles.

{\it Second.} The quantum particles (energy bundles) should
gravitate. Hence, strictly speaking, their behavior cannot be
described as a linear superposition. Therefore the ordinary second
quantization method (creation-annihilation of free particles) is
merely a good approximate scheme due to the weakness of gravity.
Thereby the creation and annihilation of particles are time
consuming dynamical non-linear processes. So, linear operators of
creation and annihilation (in Dirac sense) do exist as approximate
quantities.

{\it Third.} For sure there is an energy quantization but the
dynamical nature of this process is unknown. Avoiding the vacuum
stability problem, its self-energy, etc., we primary quantize,
however, the action, not energy. The relative (local) vacuum of some
problem is not the state with minimal energy, it is a state with an
extremal of some action functional.

POSTULATE 1.

\noindent {\it I assume that there are elementary quantum states
(EQS) $|\hbar a>, a=0,1,...$ of abstract Planck's oscillator whose
states correspond to the quantum motions with given Planck's action
quanta. Thereby only action subject to primary quantization but the
quantization of dynamical variables such as energy, spin, etc.,
postponed to dynamical stage. Physical oscillators are distributed
due to spice-time dependence of frequencies obeying field equations
which should established due to a new variation problem.}

Presumably there are some non-linear field equations whose
soliton-like solution provides the quantization of the dynamical
variables but their field carriers are smeared in dynamical
space-time. Therefore, quantum ``particles'', and, hence, their
numbers should arise as some countable solutions of non-linear wave
equations. In order to establish acceptable field equation capable
intrinsically describe all possible degrees of freedom defreezing
under intensive interaction we should to build some {\it universal
ambient Hilbert state space} $\cal{H}$. I will use {\it the
universality of the action} whose variation capable generate any
dynamical variable. Vectors of {\it action state space} $\cal{H}$ I
will call {\it action amplitude} (AA). Some of them will be EQS's of
motion corresponding to entire numbers of Planck's quanta $| \hbar
a>$. Generally (AA) are their coherent superposition
\begin{eqnarray}
|G>=\sum_{a=0}^{\infty} g^a| \hbar a>.
\end{eqnarray}
may represented of the ground state of some quantum system. In order
to avoid the misleading reminiscence about Schr\"odinger state
vector I will use $|G>,|S>$, instead of $|\Psi>$. Since the action
in itself does not create gravity, it is legible to create such
linear superposition of $|\hbar a>=(a!)^{-1/2} ({\hat \eta^+})^a|0>$
constituting $SU(\infty)$ multiplete of the Planck's action quanta
operator $\hat{S}=\hbar {\hat \eta^+} {\hat \eta}$ with the spectrum
$S_a=\hbar a$ in the separable Hilbert space $\cal{H}$. The standard
basis $\{|\hbar a>\}_0^{\infty}$ will be used with the `principle'
quantum number $a=0,1,2...$ assigned by Planck's quanta counting.

Since any ray AA has isotropy group $H=U(1)\times U(N)$, in
$\cal{H}$ effectively act only coset transformations
$G/H=SU(N)/S[U(1) \times U(N-1)]=CP(N-1)$. Therefore the ray
representation of $SU(N)$ in $C^N$, in particular, the embedding of
$H$ and $G/H$ in $G$, is the state-depending parametrization.
Therefore, there is a diffeomorphism between space of the rays
marked by the local coordinates in the map $U_j:\{|G>,|g^j| \neq 0
\}, j>0$
\begin{eqnarray}\label{coor}
\pi^1=\frac{g^0}{g^j},...,\pi^j=\frac{g^{j-1}}{g^j},
\pi^{j+1}=\frac{g^{j+1}}{g^j},..., \pi^{N-1}=\frac{g^{N-1}}{g^j},...
\end{eqnarray}
and the group manifold of the coset transformations
$G/H=SU(N)/S[U(1) \times U(N-1)]=CP(N-1)$ \cite{Besse}, where $N \to
\infty $. This diffeomorphism provided by the coefficient functions
$\Phi^i_{\alpha}$ of the local generators (see below). The choice of
the map $U_j$ means, that the comparison of quantum amplitudes
refers to the amplitude with the action $\hbar j$. The breakdown of
$SU(\infty)$ symmetry on each AA to the isotropy group $H=U(1)\times
U(\infty)$ contracts full dynamics down to $CP(\infty)$. The
physical interpretation of these transformations is given by the

POSTULATE 2.

\noindent {\it Super-equivalence  principle: the unitary
transformations of the AA may be identified with the physical
unitary fields.  The coset transformation
$G/H=SU(\infty)/S[U(1)\times U(\infty)]=CP(\infty)$ is the quantum
analog of classical force: its action is equivalent to some
physically distinguishable variation of AA in $CP(\infty)$}.

I will assume that all ``vacua'' solutions belong to single
separable {\it projective Hilbert space} $CP(N-1)$. The vacuum is
now merely the stationary point of some action functional, not
solution with the minimal energy. Energy will be associated with
tangent vector field to $CP(N-1)$ giving velocity of the action
variation in respect with a ``second time'' \cite{Prig} close to the
notion of Newton-Stueckelberg-Horwitz-Piron (NSHP) time \cite{H1}.
Dynamical space-time will be built at any vacuum (see below).
Therefore Minkowskyan space-time is functionally local in $CP(N-1)$
and the space-time motion dictated by the field equations connected
with two infinitesimally close ``vacua''. The connection between
these local space-times may be physically established by the
measurement given in terms of geometry of the base manifold
$CP(N-1)$. It seems like the Everett's idea about ``parallel
words'', but has of course different physical sense. Now we are
evidences of the Multiverse concept \cite{W1}. I think there is only
one Universe but there exists continuum of dynamical space-times
each of them related to one point of the ``vacuum landscape''
$CP(N-1)$. The standard approach, identifying Universe with
space-time, is too strong assumption from this point of view.

\section{Local dynamical variables during NSHP
evolution} When quantum setup traverses in $CP(\infty)$, the
evolution curve may be associated with trace of the vacuum extremal
in $CP(\infty)$ under the action variation $S+\delta S$. The length
of the evolution curve in $CP(\infty)$ may be measured in seconds.
Then the length of the evolution curve may be identified with the
NSHP time. The velocities (rates of change the action against NSHP
time) of transition from one superposition state to another is a
measure of system energy. These velocities are tangent vectors to
$CP(N-1)$. On the other hand they are operators of differentiation
(variation) of some functionals (mathematical scalar fields)
$\mathcal{D}S=F^i(\Omega)\frac{\partial S}{\partial \pi^i} + c.c.$
or operators (mathematical vector fields)
$\mathcal{D}V^n=F^i(\Omega)(\frac{\partial V^n}{\partial \pi^i}
+\Gamma^n_{ik}V^k) + c.c.$ over $CP(\infty)$. The distribution of
frequencies (energies) will be established by some non-linear field
equation which arise as the condition of the parallel transport of
the Hamiltonian vector field \cite{Le1,Le2}.

In order to build Hamiltonian vector field one needs a convenient
representation of the hermitian Hamiltonian. It is well known that
each hermitian matrix may be represented as the linear combination
of Habbard matrix $\hat{B^i_k}$
\begin{eqnarray}
\hat{H}=E^k_i \hat{B^i_k}=E^k_i\left(%
\begin{array}{ccccccccc}
  0 & 0 & . & . & . & 0 & . & . & . \\
  0 & 0 & . & . & . & 0 & . & . & . \\
  . & . & . & . & . & . & . & . & . \\
  . & . & . & . & . & . & . & . & . \\
  . & . & . & . & . & 0 & . & . & . \\
  0 & 0 & . & . & 0 & 1 & 0 & . & . \\
  . & . & . & . & . & 0 & . & . & . \\
  . & . & . & . & . & . & . & . & . \\
  . & . & . & . & . & . & . & . & . \\
\end{array}%
\right),
\end{eqnarray}
where $E^k_i=E^{i*}_k$. Avoiding two-index numeration let me use
$SU(N)$ traceless generators like Pauli, Gell-Mann, etc., matrices.
The problem is to find $\hat{\Lambda_{\alpha}}$ for given
$\hat{B^i_k}$ and the inverse problem. For any $N \geq 2$ one has
the follows numeration scheme
\begin{eqnarray}
for \quad n=2; \hat{\Lambda_1}=\hat{B^1_2} + \hat{B^2_1};
\hat{\Lambda_2}=i(\hat{B^2_1} - \hat{B^1_2}); \cr
\hat{\Lambda_3}=\frac{1}{\sqrt{1}}(\hat{B^1_1} - \hat{B^2_2}); \cr
for \quad n=3; \hat{\Lambda_4}=\hat{B^1_3} + \hat{B^3_1};
\hat{\Lambda_5}=i(\hat{B^1_3} - \hat{B^3_1}); \cr
\hat{\Lambda_6}=\hat{B^2_3} + \hat{B^3_2};
\hat{\Lambda_7}=i(\hat{B^2_3} - \hat{B^3_2}); \cr
\hat{\Lambda_8}=\frac{1}{\sqrt{3}}(\hat{B^1_1}+\hat{B^2_2} - 2
\hat{B^2_2}); \cr for \quad n=4; \hat{\Lambda_9}=\hat{B^1_4} +
\hat{B^4_1}; \hat{\Lambda}_{10}=i(\hat{B^1_4} - \hat{B^4_1}); \cr
\hat{\Lambda_{11}}=\hat{B^2_4} + \hat{B^4_2};
\hat{\Lambda_{12}}=i(\hat{B^2_4} - \hat{B^4_2}); \cr
\hat{\Lambda_{13}}=(\hat{B^3_4} + \hat{B^4_3});
\hat{\Lambda_{14}}=i(\hat{B^3_4} - \hat{B^4_3}); \cr
\hat{\Lambda_{15}}=\frac{1}{\sqrt{6}}(\hat{B^1_1}+\hat{B^2_2} +
\hat{B^3_3} - 3\hat{B^4_4}); \cr . \cr .\cr. \cr
\end{eqnarray}
Let me introduce $m=min(i,k) \geq 1$ and $M=max(i,k) \geq 2$. Then
it is clear that $\hat{\Lambda}_{\alpha}$ belongs to the set of the
$2(M-1)+1$ matrices
\begin{eqnarray}{\label{set1}}
for \quad n=M; \hat{\Lambda}_{M^2-2M+1}=\hat{B}^1_{M-1} +
\hat{B}^{M-1}_1; \hat{\Lambda}_{M^2-2M+2}=i(\hat{B}^1_{M-1} -
\hat{B}^{M-1}_1);  \cr . \cr . \cr . \cr
\hat{\Lambda}_{M^2-2M+2m-1}=\hat{B}^m_{M-1} + \hat{B}^{M-1}_m;
\hat{\Lambda}_{M^2-2M+2m}=i(\hat{B}^m_{M-1} - \hat{B}^{M-1}_m) \cr .
\cr . \cr . \cr
\hat{\Lambda}_{M^2-1}=\sqrt{\frac{2}{M(M-1)}}(\hat{B^1_1}+\hat{B^2_2}
+ \hat{B^3_3}+...+\hat{B}^{M-1}_{M-1} - (M-1)\hat{B^M_M});
\end{eqnarray}
Therefore two matrices $\hat{B}^i_k$ and $\hat{B}^k_i$ if $i \neq k$
define the matrix $\hat{\Lambda}_{M^2-2M+2m-1}=\hat{B}^m_{M-1} +
\hat{B}^{M-1}_m$ and matrix
$\hat{\Lambda}_{M^2-2M+2m}=i(\hat{B}^m_{M-1} - \hat{B}^{M-1}_m) $,
but if $i=k$, then $\hat{B}^M_M$  corresponds to the diagonal matrix
$\hat{\Lambda}_{M^2-1}=\sqrt{\frac{2}{M(M-1)}}(\hat{B^1_1}+\hat{B^2_2}
+ \hat{B^3_3}+...+\hat{B}^{M-1}_{M-1} - (M-1)\hat{B^M_M}) $.

In order to solve the inverse problem, namely, for given
$\hat{\Lambda}_{\alpha} $ to find the correspond matrix
$\hat{B}^k_i$ one should to find maximal solution of the inequality
$M^2 \leq \alpha +1$. It means that $\hat{\Lambda}_{\alpha} $
belongs to the set (\ref{set1}). If some entire $M$ is such that
$M^2=\alpha+1$, then one has the diagonal matrix $\hat{B}^M_M$. But
if $M^2 < \alpha +1$ then it is easy  to find the position of given
$\hat{\Lambda}_{\alpha} $ from the set (\ref{set1})t. Namely, the
quantity of rows $M-1$ in the set gives us the index $i=M-1$, and
the number of row containing $\hat{\Lambda}_{\alpha} $ gives us the
index $k=m$ from one of the equations
\begin{eqnarray}{\label{m}}
m=\frac{\alpha-M^2+2M-1}{2} \quad or \quad
m=\frac{\alpha-M^2+2M}{2},
\end{eqnarray}
depending on which nominator is even. Therefore any traceless
hermitian operator having matrix representation may be represented
as a single index linear combination of the $\hat{\Lambda}_{\alpha}
$ matrices. For example, the operator of harmonic oscillator
coordinate is as follows
\begin{eqnarray}{\label{oscX}}
\hat{x} = \sqrt{\frac{\hbar}{2 m \omega}} \sum_{n=1}^\infty
\sqrt{n}\hat{\Lambda}_{(n+1)^2-3},
\end{eqnarray}
and for the momentum operator one has
\begin{eqnarray}{\label{oscP}}
\hat{p} = \sqrt{\frac{2 m \omega}{\hbar}} \sum_{n=1}^\infty
\sqrt{n}\hat{\Lambda}_{(n+1)^2-2}.
\end{eqnarray}
Since commutators and anti-commutators of $\hat{\Lambda}_{\alpha}$
matrices give linear combination of the unit matrix and
$\hat{\Lambda}_{\alpha}$ matrices, such algebraic operations as
summation and multiplication lead to algebra of the unitary group
$Alg U(\infty)$. Hence many Hermitian reasonable Hamiltonian
$\hat{H}$ may be represented as follows:
\begin{eqnarray}{\label{abstrH}}
\hat{H} = \epsilon \hat{\Delta} + \hbar \sum_{n=1}^\infty
\Omega^{\alpha} \hat{\Lambda}_{\alpha},
\end{eqnarray}
where $\hat{\Delta}$ is unit matrix. Then $\hbar \sum_{n=1}^\infty
\Omega^{\alpha} \hat{\Lambda}_{\alpha}=\hat{H} - \epsilon
\hat{\Delta}$. Multiplying both sides by $\hat{\Lambda}_{\beta}$ and
taking into account $Tr(\hat{\Lambda}_{\beta}
\hat{\Lambda}_{\alpha})= 2\delta_{\alpha \beta}$, we get
\begin{eqnarray}{\label{Tr}}
Tr(\hat{\Lambda}_{\beta}(\hat{H} - \epsilon \hat{\Delta})) =\hbar
\sum_{n=1}^\infty \Omega^{\alpha} Tr(\hat{\Lambda}_{\beta}
\hat{\Lambda}_{\alpha})= 2\hbar \Omega^{\beta} ,
\end{eqnarray}
and, therefore, the ``multipole'' coefficient functions may be
extracted
\begin{eqnarray}{\label{Omega}}
\Omega^{\alpha} = \frac{1}{2\hbar} Tr(\hat{\Lambda}_{\alpha}(\hat{H}
- \epsilon \hat{\Delta}))=\frac{1}{2\hbar}
Tr(\hat{\Lambda}_{\alpha}\hat{H}).
\end{eqnarray}
The coefficients $\Omega^{\alpha}$ for the Hamiltonian of the
harmonic oscillator
\begin{eqnarray}
\hat{H}_{HO}=\hbar \omega \left(%
\begin{array}{ccccccccc}
  1/2 & 0 & . & . & . & 0 & . & . & . \\
  0 & 3/2 & . & . & . & 0 & . & . & . \\
  . & . & . & . & . & . & . & . & . \\
  . & . & . & . & . & . & . & . & . \\
  . & . & . & . & . & 0 & . & . & . \\
  0 & 0 & . & . & 0 & n+1/2 & 0 & . & . \\
  . & . & . & . & . & 0 & . & . & . \\
  . & . & . & . & . & . & . & . & . \\
  . & . & . & . & . & . & . & . & . \\
\end{array}%
\right),
\end{eqnarray}
is for off diagonal matrices as follows:
\begin{eqnarray}{\label{OmegaND}}
\Omega^{\alpha} = \frac{1}{2\hbar}
Tr(\hat{\Lambda}_{\alpha}\hat{H}_{HO})= \omega(m-1/2+M-3/2)/2=
\omega (m+M-2)/2
\end{eqnarray}
and for the diagonal matrices $\hat{\Lambda}_{\alpha}$ they are
given by
\begin{eqnarray}{\label{OmegaD}}
\Omega^{\alpha} = \frac{1}{2\hbar}
Tr(\hat{\Lambda}_{\alpha}\hat{H}_{HO})= \omega
 \sqrt{\frac{1}{8M(M-1)}}(\sum_{k=1}^{M-1}k-M+1).
\end{eqnarray}
Hermitian Hamiltonian in the form (\ref{abstrH}) used in wide
spectrum of physical problems \cite{Perelomov}. In these problems
coefficient functions $\epsilon$ and $\Omega^{\alpha}$ usually arise
from the classical space-time model. In our case we should find
equations for these functions starting from conditions applied to
the LDV represented by vector fields in functional space. In other
words one should find ``field shell'' of quantum particles in
dynamical space-time. Now we should introduce the dynamical
space-time based on the quantum measurements in terms of $CP(N-1)$
geometry \cite{Le1,Le2}.

Hereafter I will use the representation of the Hamiltonian vector
fields as follows:
\begin{eqnarray}{\label{HVF}}
\frac{d \pi^i}{d \tau}=H^i =\hbar \Phi^i_{\alpha} \Omega^{\alpha}=
\frac{1}{2} \Phi^i_{\alpha} Tr(\hat{\Lambda}_{\alpha}\hat{H}),
\end{eqnarray}
where
\begin{equation}{\label{Phi}}
\Phi_{\sigma}^i = \lim_{\epsilon \to 0} \epsilon^{-1}
\biggl\{\frac{[\exp(i\epsilon \lambda_{\sigma})]_m^i g^m}{[\exp(i
\epsilon \lambda_{\sigma})]_m^j g^m }-\frac{g^i}{g^j} \biggr\}=
\lim_{\epsilon \to 0} \epsilon^{-1} \{ \pi^i(\epsilon
\lambda_{\sigma}) -\pi^i \},
\end{equation}
\cite{Le3}.

Perfectly isolated quantum systems such as ``elementary'' particles
do not exist. But it is natural to assume that the elementary
excitations of the ground (vacuum) should exist as some entity,
invariant relative the choice of different physical field
constituting a setup. On the other hand one should avoid the
artificial distinction between the bare ``elementary'' particle and
its ``surrounding'' field.

In the framework of $SU(N)$ symmetry I use a model of the quantum
field theory based on a {\it local Hamiltonian of interaction}. It
consists of the sum of $N^2-1$ the energies of the `elementary
systems' (particle plus fields) is equal to the excitation energy of
the GCS, and the local Hamiltonian $\vec{H}$ is linear against the
partial derivatives $\frac{\partial }{\partial \pi^i} = \frac{1}{2}
(\frac{\partial } {\partial \Re{\pi^i}} - i \frac{\partial
}{\partial \Im{\pi^i}})$ and $\frac{\partial }{\partial \pi^{*i}} =
\frac{1}{2} (\frac{\partial } {\partial \Re{\pi^i}} + i
\frac{\partial }{\partial \Im{\pi^i}})$, i.e. it is the tangent
vector to $CP(N-1)$
\begin{eqnarray}\label{field}
\vec{H}= \hbar
\Omega^{\alpha}\Phi^i_{\alpha}\frac{\partial}{\partial \pi^i} + c.c.
\cr = \vec{T_h}+\vec{U_b} = \hbar \Omega^h \Phi_h^i \frac{\partial
}{\partial \pi^i} + \hbar \Omega^b \Phi_b^i \frac{\partial
}{\partial \pi^i} + c.c..
\end{eqnarray}
This Hamiltonian describes the interaction between quantum system,
given be the local analog of the spin-operators (Pauli, Gell-Mann,
etc.) expressed in coordinates $(\pi^1,...,\pi^i,...)$, and its
adjoint unitary ``field shell'' $\Omega^{\alpha} $ actually
transforming states of the quantum system. We have in fact a
self-consistent problem but under some reasonable assumptions this
may be reduced to pure non-linear field equations for the ``field
shell''.

The dynamical variables corresponding symmetries of the GCS and
their breakdown should be expressed now in terms of the local
coordinates $\pi^k$. Hence the internal dynamical variables and
their norms should be state-dependent, i.e. local in the state space
\cite{Le1,Le2,Le3,Le4}. These local dynamical variables realize a
non-linear representation of the unitary global $SU(N)$ group in the
Hilbert state space $C^N$. Namely, $N^2-1$ generators of $G = SU(N)$
may be divided in accordance with Cartan decomposition: $[B,B] \in
H, [B,H] \in B, [B,B] \in H$. The $(N-1)^2$ generators
\begin{eqnarray} \Phi_h^i \frac{\partial}{\partial \pi^i}+c.c. \in
H,\quad 1 \le h \le (N-1)^2
\end{eqnarray}
 of the isotropy group $H = U(1)\times
U(N-1)$ of the ray (Cartan sub-algebra) and $2(N-1)$ generators
\begin{eqnarray}
\Phi_b^i \frac{\partial}{\partial \pi^i} + c.c. \in B, \quad 1 \le b
\le 2(N-1)
\end{eqnarray}
are the coset $G/H = SU(N)/S[U(1) \times U(N-1)]$ generators
realizing the breakdown of the $G = SU(N)$ symmetry of the GCS.
Furthermore, $(N-1)^2$ generators of the Cartan sub-algebra may be
divided into the two sets of operators: $1 \le c \le N-1$ ($N-1$ is
the rank of $Alg SU(N)$) Abelian operators, and $1 \le q \le
(N-1)(N-2)$ non-Abelian operators corresponding to the
non-commutative part of the Cartan sub-algebra of the isotropy
(gauge) group. Here $\Phi^i_{\sigma}, \quad 1 \le \sigma \le N^2-1 $
are the coefficient functions of the generators of the non-linear
$SU(N)$ realization. They give the infinitesimal shift of
$i$-component of the coherent state driven by the $\sigma$-component
of the unitary multipole field rotating the generators of $Alg
SU(N)$ and they are defined by the formulas (\ref{Phi}). These show
how look $SU(N)$ generators in the vicinity of the origin of the map
$U_j:|G>, |g^j| \neq 0$ being expressed in the local ray's
coordinates $(\pi^1,...,\pi^i,...,\pi^{N-1})$. This representation
is useful since embedding Cartan sub-group, coset manifold, etc.,
are state-dependent, i.e. requires local coordinates.

Now I will introduce the Lagrangian $\Lom$ and the canonical
momentum $P_i$. Let me put
\begin{equation} \label{2}
\Lom=||E||=\hbar
\sqrt{G_{ik^*}(\Omega^{\alpha}\Phi^i_{\alpha})(\Omega^{\beta}
\Phi^{k}_{\beta})^*}=\hbar \sqrt{G_{ik^*}V^i V^{k*}}
\end{equation}
since
\begin{equation}\label{3}
V^i=\frac{d \pi^i}{d\tau}=\Omega^{\alpha}\Phi^i_{\alpha}.
\end{equation}
Then the canonical momentum is as follows
\begin{equation}\label{4}
P_i= \frac{\partial \Lom}{\partial V^i}= \frac{\hbar^2}{2 ||E||}
G_{ik^*}V^{k*}
\end{equation}
and its Hermitian conjugated value is
\begin{equation}\label{5}
P^{i*}= \frac{\hbar^2}{2 ||E||} G^{i*k}
G_{ks^*}V^{s*}=\frac{\hbar^2}{2 ||E||}V^{i*}.
\end{equation}
Therefore,
\begin{equation}\label{6}
P^{i}= \frac{\hbar^2}{2 ||E||}V^{i},
\end{equation}
and then the generator of the momentum (tangent vector) is as
follows
\begin{equation}\label{7}
\vec{P}= P^{i}\frac{\partial}{\partial \pi^i} + c.c. =
\frac{\hbar^2}{4 ||E||}V^{i}\frac{\partial}{\partial \pi^i} + c.c.
\end{equation}
Since the velocity $V^{i}$ has physical dimension of frequency, the
momentum $P^{i}$ has the physical dimension of the action. The
contraction of corresponding LDV's has the square of modulus
\begin{equation}\label{8}
P^{i}P_i= \frac{\hbar^4}{4
||E||^2}G_{ik*}V^{i}V^{k*}=\frac{\hbar^2}{4}
\end{equation}
equal to the minimal uncertainty.
\section{Dynamical quantum space-time}
The internal hidden dynamics of the quantum  configuration given by
AA should be somehow reflected in physical space-time. Therefore we
should solve the ``inverse representation problem'': to find locally
unitary representation of dynamical group $SU(N)$ in the dynamical
space-time where acts the induced realization of the coherence group
$SU(2)$ of the Qubit spinor \cite{Le1,Le2}. Its components subjected
to the ``quantum Lorentz transformations'' \cite{G}. We should build
the local spinor basis invariantly related to the ground states
manifold $CP(N-1)$. First of all we have to have the local reference
frame (LRF) as some analog of the ``representation'' of $SU(N)$.
Each LRF and, hence, $SU(N)$ ``representation'' may be marked by the
local coordinates (\ref{coor}) of the ``vacuum landscape''. Now we
should almost literally repeat differential geometry of a smooth
manifold embedded in flat ambient Hilbert space. The geometry of
this smooth manifold is the projective Hilbert space equipped with
the Fubini-Study metric that may be expressed in the local
coordinates as follows
\begin{equation}\label{FS}
G_{ik^*} = \frac{(1+\sum |\pi^s|^2) \delta_{ik}-\pi^{i^*} \pi^k}
{(1+\sum |\pi^s|^2)^2},
\end{equation}
and with the affine connection
\begin{eqnarray}\label{Gamma}
\Gamma^i_{mn} = \frac{1}{2}G^{ip^*} (\frac{\partial
G_{mp^*}}{\partial \pi^n} + \frac{\partial G_{p^*n}}{\partial
\pi^m}) = - \frac{\delta^i_m \pi^{n^*} + \delta^i_n
\pi^{m^*}}{1+\sum |\pi^s|^2}.
\end{eqnarray}

The velocity of ground state evolution relative NSHP time is given
by the formula
\begin{eqnarray}\label{41}
|H> = \frac{d|G>}{d\tau}=\frac{\partial g^a}{\partial
\pi^i}\frac{d\pi^i}{d\tau}|a\hbar>=|T_i>\frac{d\pi^i}{d\tau}=H^i|T_i>,
\end{eqnarray}
 is the tangent vector to the evolution curve
$\pi^i=\pi^i(\tau)$, where
\begin{eqnarray}\label{42}
|T_i> = \frac{\partial g^a}{\partial \pi^i}|a\hbar>=T^a_i|a\hbar>.
\end{eqnarray}
Then the ``acceleration'' is as follows
\begin{eqnarray}\label{43}
|A> =
\frac{d^2|G>}{d\tau^2}=|g_{ik}>\frac{d\pi^i}{d\tau}\frac{d\pi^k}{d\tau}
+|T_i>\frac{d^2\pi^i}{d\tau^2}=|N_{ik}>\frac{d\pi^i}{d\tau}\frac{d\pi^k}{d\tau}\cr
+(\frac{d^2\pi^s}{d\tau^2}+\Gamma_{ik}^s
\frac{d\pi^i}{d\tau}\frac{d\pi^k}{d\tau})|T_s>,
\end{eqnarray}
where
\begin{eqnarray}\label{44}
|g_{ik}>=\frac{\partial^2 g^a}{\partial \pi^i \partial \pi^k}
|a\hbar>=|N_{ik}>+\Gamma_{ik}^s|T_s>
\end{eqnarray}
and the normal state
\begin{eqnarray}\label{45}
|N> = N^a|a\hbar>=(\frac{\partial^2 g^a}{\partial \pi^i \partial
\pi^k}-\Gamma_{ik}^s \frac{\partial g^a}{\partial \pi^s})
\frac{d\pi^i}{d\tau}\frac{d\pi^k}{d\tau}|a\hbar>
\end{eqnarray}
to the ``hypersurface'' of the ground states. Then the minimization
of this ``acceleration'' under the transition from point $\tau$ to
$\tau+d\tau$ may be achieved by the annihilation of the tangential
component
\begin{equation}
(\frac{d^2\pi^s}{d\tau^2}+\Gamma_{ik}^s
\frac{d\pi^i}{d\tau}\frac{d\pi^k}{d\tau})|T_s>=0
\end{equation}
i.e. under the condition of the affine parallel transport of the
Hamiltonian vector field
\begin{equation}\label{par_tr}
dH^s +\Gamma^s_{ik}H^id\pi^k =0.
\end{equation}

The Gauss-Codazzi equations
\begin{eqnarray}\label{46}
\frac{\partial N^a}{\partial \pi^i}=B^s_i T^a_s \cr \frac{\partial
T_k^a}{\partial \pi^i}-\Gamma^s_{ik}T^a_s=B_{ik}N^a
\end{eqnarray}
I used here instead of the anthropic principle \cite{W1}. These give
us dynamics of the vacuum (normal) vector and the tangent vectors,
i.e. one has the LRF dynamics modeling the ``moving representation''
or moving quantum setup
\begin{eqnarray}\label{47}
\frac{d N^a}{d \tau}= \frac{\partial N^a}{\partial \pi^i} \frac{d
\pi^i}{d\tau}+c.c.=B^s_i T^a_s \frac{d \pi^i}{d\tau} +c.c. = B^s_i
T^a_s H^i +c.c. \cr \frac{d T_k^a}{d \tau} =\frac{\partial
T_k^a}{\partial \pi^i}\frac{d \pi^i}{d\tau} +c.c.=
  (B_{ik}N^a+\Gamma^s_{ik}T^a_s)\frac{d \pi^i}{d\tau}+c.c.
= (B_{ik}N^a+\Gamma^s_{ik}T^a_s) H^i+c.c.
\end{eqnarray}
Please, remember that $0 \leq a \leq N$, but $1\leq i,k,m,n,s \leq
N-1$. In order to find the matrix $B^s_i$ of the second quadratic
form of the ground states ``hypersurface'' we shall use the
self-adjoint Weingarten mapping
$L_{(\pi^1,...,\pi^{N-1})}(\vec{D})=-\nabla_{\vec{D}}|N>$
\cite{Thorp} where $\vec{D}$ refers to some local dynamical variable
(LDV) \cite{Le5}.

If we would like to have some embedding of the ``Hilbert (quantum)
dynamics" in space-time we should to formalize the quantum
observation (or measurement of some dynamical variable).

The measurement, i.e. attributing a number to some dynamical
variable or observable has in physics subjective as well as
objective sense. Namely: the numeric value of some observable
depends as a rule on a setup (the character of  motion of
laboratory, type of the measuring device, field strength, etc.).
However the relationships between numeric values of dynamical
variables and numeric characteristics of laboratory motion, field
strength, etc., should be formulated as invariant, since they
reflect the objective character of the physical interaction used in
the measurement process. The numbers obtained  due to the
measurements carry information which does not exist a priori, i.e.
before the measurement process. But the information comprised of
subjective as well as objective invariant part reflects the physics
of interaction. The last is one of the main topics of  quantum field
theory. Since each measurement reducible (even if it is unconscious)
to the answer of the question ''yes'' or ''no'', it is possible to
introduce formally a quantum dynamical variable "logical spin 1/2"
\cite{Le3} whose coherent states represent the quantum bit of
information "Qubit".

\pagebreak

POSTULATE 3

{\it I assume that the invariant i.e. physically essential part of
information represented by the coherent states of the "logical spin
1/2" is related to the space-time structure.}

Such assumption is based on the observation that on one side the
space-time is the manifold of points artificially depleted of all
physical characteristics (material points without reference to
masses). In principle arbitrary local coordinates may be attributed
to these points. On the other hand as we know from general
relativity the metric structure depends on the matter distribution
and the zero approximation of the metric tensor $g_{\mu
\nu}=\eta_{\mu \nu}+...$ gives the Lorentz invariant interval. The
spinor structure Lorentz transformations represents the
transformations of the coherent states of the "logical spin 1/2" or
"Qubit". Thereby we can assume the measurement of the quantum
dynamical variables expressed by the "Qubit" spinor ``creates'' the
local space-time coordinates. I will formulate non-linear field
equations in this local space-time due to a variational principle
referring to the generator of the quantum state deformation.

Now one should build the Qubit spinor in the local basis $(|N>,|D>)$
for the quantum question with ``yes'' or ``no'' spectrum in respect
with the measurement of some local dynamical variable $\vec{D}$. I
will assume that there is {\it natural state $|D>$ of the quantum
system in the LRF representation}  equal to the lift of LDV $\vec{D}
\in T_{\pi}CP(N-1)$ into the environmental Hilbert space $\cal{H}$,
and there is {\it expectation state}
$|D_{expect}>=\alpha_0|N>+\beta_0|\widetilde{D}>$, associated with
the ``measuring device'' tuning. This notional measuring device is
associate with the local unitary projector along the normal $|N>$
and onto the natural state $\widetilde{|D>}$. In fact it defines the
covariant derivative in $CP(N-1)$. The lift-vectors $|N>,|D>$ are
given by the solutions of (\ref{47}) arising under the measurement
of the LDV $\vec{D}$. In general $|D>$ it is not a tangent vector to
$CP(N-1)$. But renormalized vector defined as the covariant
derivative $|\widetilde{D}>=|D>-<Norm|D>|Norm>$ is a tangent vector
to $CP(N-1)$ if $|Norm>=\frac{|N>}{\sqrt{<N|N>}}$. The operation of
the velocity renormalization is the orthogonal (unitary) projector.
Indeed,
\begin{eqnarray}
\widetilde{|\widetilde{D}>}= \widetilde{|D>-<Norm|D>|Norm>}\cr =
|D>-<Norm|D>|Norm> \cr - <Norm|(|D>-<Norm|D>|Norm>)|Norm> \cr
=|D>-<Norm|D>|Norm> = |\widetilde{D}>.
\end{eqnarray}
Then at the point $(\pi^1,...,\pi^{N-1})$ one has two components of
the Qubit spinor
\begin{eqnarray}\label{513}
\alpha_{(\pi^1,...,\pi^{N-1})}=\frac{<N|D_{expect}>}{<N|N>} \cr
\beta_{(\pi^1,...,\pi^{N-1})}=\frac{<\widetilde{D}|D_{expect}>}
{<\widetilde{D}|\widetilde{D}>}
\end{eqnarray}
then at the infinitesimally close point
$(\pi^1+\delta^1,...,\pi^{N-1}+\delta^{N-1})$ one has new Qubit
spinor
\begin{eqnarray}\label{514}
\alpha_{(\pi^1+\delta^1,...,\pi^{N-1}+\delta^{N-1})}=\frac{<N'|D_{expect}>}
{<N'|N'>} \cr \beta_{(\pi^1+\delta^1,...,\pi^{N-1}+\delta^{N-1})}=
\frac{<\widetilde{D}'|D_{expect}>}{<\widetilde{D}'|\widetilde{D}'>}
\end{eqnarray}
where the basis $(|N'>,|\widetilde{D}'>)$ is the lift of the
parallel transported $(|N>,|\widetilde{D}>)$ from the
infinitesimally close $(\pi^1+\delta^1,...,\pi^{N-1}+\delta^{N-1})$
back to $(\pi^1,...,\pi^{N-1})$.

These two spinors being connected with infinitesimal ``Lorentz spin
transformations matrix'' \cite{G} $L$ create the local dynamical
space-time \cite{Le1,Le2}. The coordinates $x^\mu$ of points in this
space-time serve in fact merely for parametrization of deformations
of the ``field shell'' arising under its motion according to
non-linear field equations.

\section{Nonlinear gauge field equations for transition from
infinitesimally close dynamical quantum space-times} ``Particle''
now associated with its ``field shell'' in the dynamical space-time.
At each GCS $(\pi^1,...,\pi^{N-1})$ of the $CP(N-1)$ one has an
``expectation value'' of LDV's defined by a measuring device. But
this GCS may by reached along of one of continuum pathes. Therefore
the comparison of two vector fields and their ``expectation values''
in neighborhood points requires some natural rule. The ``natural''
in our case means that the comparison has sense only for same
``field shell''. For this reason one should have a
``self-identification'' procedure. The affine parallel transport in
$CP(N-1)$ of vector fields is a natural and the simplest rule for
the comparison of generators corresponding ``field shells''.
Physically the self-identification of ``particle'' at different GCS,
i.e. setups, literally means that vector fields corresponding to its
LDV's are covariant constants relative the Fubini-Study metric.
Since we have only the unitary fields as parameters of GCS
transformations I assume that in accordance with the
super-equivalence principle under the infinitesimal shift of the
unitary field $\delta \Omega^{\alpha}$ in the dynamical space-time,
the shifted Hamiltonian field should coincide with the infinitesimal
shift of tangent Hamiltonian field generated by the parallel
transport in $CP(N-1)$  during NSHP time $\delta \tau)$. Hence,
under small specific variation of the unitary fields
$\Omega^{\alpha}$, obeying field equations given below, two
dynamical variables
\begin{eqnarray} \label{ym}
\mathcal{\vec{D}}_1=F_1^i(\pi, \Omega)\frac{\partial}{\partial
\pi^i} + c.c.,\cr \mathcal{\vec{D}}_2=F_2^i(\pi,
\Omega)\frac{\partial}{\partial \pi^i} + c.c.
\end{eqnarray}
such as Hamiltonian $H^i$ and momentum $P^i$
\begin{eqnarray} \label{EM}
F_1^i(\pi, \Omega)=H^i=\hbar\Omega^{\alpha}\Phi^i_{\alpha} =\hbar
V^i \cr F_2^i(\pi, \Omega)=P^i= \frac{\hbar^2}{2 ||E||}V^{i},
\end{eqnarray}
subjected to the infinitesimal parallel transport. Then assuming
that the affine parallel transport of each dynamical variable
generated by the Hamiltonian along the evolution curve should be
accompanied with specific variations of $\Omega^\alpha$ one has
\begin{eqnarray} \label{var} F_{1,2}^i(\pi, \Omega+\delta
\Omega)=F_{1,2}^i(\pi, \Omega)-\Gamma ^i_{mn} F^n_{1,2}(\pi, \Omega)
V^m \delta \tau,
\end{eqnarray}
or separately for $H^i$ and $P^i$:
\begin{eqnarray}\label{51}
\hbar (\Omega^{\alpha} + \delta \Omega^{\alpha} ) \Phi^k_{\alpha}=
\hbar \Omega^{\alpha}( \Phi^k_{\alpha} - \Gamma^k_{mn}
\Phi^m_{\alpha} V^n \delta \tau), \cr P^i(\Omega^{\alpha}+\delta
\Omega^{\alpha})=P^i(\Omega^{\alpha})-\Gamma^k_{mn}P^m(\Omega^{\alpha})V^n
\delta \tau.
\end{eqnarray}
Then taking into account
\begin{eqnarray}\label{52}
P^i(\Omega+\delta \Omega,\Omega^*+\delta
\Omega^*)=P^i(\Omega,\Omega^*)+\frac{\partial
P^i(\Omega,\Omega^*)}{\partial \Omega^{\sigma}}\delta
\Omega^{\sigma}+\frac{\partial P^i(\Omega,\Omega^*)}{\partial
\Omega^{\sigma^*}}\delta \Omega^{\sigma^*}+...
\end{eqnarray}
where complex partial derivatives are as follow $\frac{\partial
}{\partial \Omega} = \frac{1}{2} (\frac{\partial } {\partial
\Re{\Omega}} - i \frac{\partial }{\partial \Im{\Omega}})$ and
$\frac{\partial }{\partial \Omega^{*}} = \frac{1}{2} (\frac{\partial
} {\partial \Re{\Omega}} + i \frac{\partial }{\partial
\Im{\Omega}})$ one has the differentials
\begin{eqnarray}\label{53}
\delta \Omega^{\alpha} &=& - \Omega^{\alpha}\Gamma^m_{mn} V^n \delta
\tau, \cr \delta P^i &=& \frac{\partial
P^i(\Omega,\Omega^*)}{\partial \Omega^{\sigma}}\delta
\Omega^{\sigma} \cr &=& \frac{\hbar}{4}(\Phi^i_{\sigma}(G_{ik^*}V^i
V^{k*})^{-1/2}-\frac{1}{2}V^i G_{jk^*} \Phi^j_{\sigma}
V^{k*}(G_{ik^*}V^i V^{k*})^{-3/2})\delta \Omega^{\sigma}.
\end{eqnarray}

Assuming that infinitesimal coordinates variation is generated by
the Poincar\'e group transformations $\delta
x^{\mu}=\Lambda^{\mu}_{\nu}x^{\nu}\delta\tau+u^{\mu} \delta s$,
where $\delta s=c \delta t$ is ``proper time'', one has
``4-velocity'' of evolution relative NSHP time
\begin{equation}\label{58}
U^{\mu}=\frac{\delta x^{\mu}}{\delta\tau}
=\Lambda^{\mu}_{\nu}x^{\nu}+(\frac{1}{\sqrt{1-v^2/c^2}},
\frac{\bf{v}}{c \sqrt{1-v^2/c^2}})\frac{\delta s}{\delta \tau}=
\Lambda^{\mu}_{\nu}x^{\nu}+cu^{\mu} \frac{\delta t}{\delta \tau},
\end{equation}
where $\Lambda^{\mu}_{\nu}$ corresponds to the
$\hat{L}=1+\frac{1}{2}\delta \tau
\vec{\sigma}(\vec{a}-i\vec{\omega})$ connecting infinitesimally
close Qubit spinors (\ref{513}),(\ref{514})  \cite{Le1,Le2}. Taking
into account
\begin{eqnarray} \label{var}
\frac{\partial F_{1,2}^i(\pi, \Omega)}{\partial \Omega^\alpha}\delta
\Omega^\alpha=\frac{\partial F_{1,2}^i(\pi, \Omega)}{\partial
\Omega^\alpha} \frac{\partial \Omega^\alpha}{\partial x^\mu}
\frac{\delta x^\mu}{\delta \tau} \delta \tau,
\end{eqnarray}
one has the two sets of field equations in the local dynamical
space-time
\begin{eqnarray} \label{FE}
\Lambda^\mu_\nu x^\nu \frac{\partial F_{1}^i(\pi, \Omega)}{\partial
\Omega^b} \frac{\partial \Omega^b}{\partial x^\mu} =-\Gamma ^i_{mn}
F^n_{1}(\pi, \Omega) V^m , \cr c u^\mu\frac{\delta t}{\delta \tau}
\frac{\partial F_{2}^i(\pi, \Omega)}{\partial \Omega^h}
\frac{\partial \Omega^h}{\partial x^\mu} =-\Gamma ^i_{mn}
F^n_{2}(\pi, \Omega) V^m.
\end{eqnarray}

The homogenous part of ``4-velocity'' is given by
$y^{\mu}=\Lambda^{\mu}_{\nu} x^{\nu} $ and, therefore, the field
equations generated by the Lorentz transformations read now as
follows:
\begin{equation}\label{KinEn}
y^{\nu} \frac{\partial \Omega^{b}}{\partial x^{\mu} } = -
\Omega^{b}\Gamma^m_{mn} V^n.
\end{equation}
Hence, the first set of $1 \leq b \leq 2(N-1)$ equations gives the
potential energy expressed by the coset components $\Omega^b$ and
compensated by homogeneous Lorentz transformations.

Let me suppose that local space-time shift (non-homogeneous part of
Poincar\'e group) should be represented by the variation of the
``field shell'' leading to the affine parallel transported LDV of
momentum. Then a straightforward calculation yields the field
quasi-linear equations in partial derivatives
\begin{eqnarray}\label{PotEn}
u^{\mu} \frac{\partial \Omega^{h}}{\partial x^{\mu}} \frac{\delta
t}{\delta \tau} \frac{c \hbar}{2}(\Phi^i_{h}(G_{ik^*}V^i
V^{k*})^{-1/2}-\frac{1}{2}V^i G_{jk^*} \Phi^j_{h} V^{k*}(G_{ik^*}V^i
V^{k*})^{-3/2}) \cr =-\Gamma^i_{mn}P^m(\Omega)V^n.
\end{eqnarray}
 Thus the second set of $1 \leq h
\leq (N-1)^2$ equations describe the kinetic energy expressed by the
isotropy group components $\Omega^h$ and compensated by the
space-time shifts.

These quasi-linear field equations in the case $N=2$ for the
spherically symmetric functions
\begin{eqnarray}\label{sol}
\Omega^1&=&\rho(t,r)\sin \Theta \cos \Phi \cr
\Omega^2&=&\rho(t,r)\sin\Theta \sin \Phi \cr \Omega^3&=&\rho(t,r)
\cos \Theta,
\end{eqnarray}
may be written as follows over-determined system of PDE's
\begin{eqnarray}\label{57}
E_1&:&C_{1t}(\rho,\Theta,\Phi)\frac{\partial \rho}{\partial
t}+C_{1r}(\rho,\Theta,\Phi)\frac{\partial \rho}{\partial
r}=R_1(\rho,\Theta,\Phi), \cr
E_2&:&C_{2t}(\rho,\Theta,\Phi)\frac{\partial \rho}{\partial
t}+C_{2r}(\rho,\Theta,\Phi)\frac{\partial \rho}{\partial
r}=R_2(\rho,\Theta,\Phi), \cr E_3&:&
C_{3t}(\rho,\Theta,\Phi,u^\mu)\frac{\partial \rho}{\partial
t}+C_{3r}(\rho,\Theta,\Phi,u^\mu)\frac{\partial \rho}{\partial
r}=R_3(\rho,\Theta,\Phi,u^\mu).
\end{eqnarray}
These field equations representing the natural ``corpuscular-wave
duality'', since their equations of characteristics
\begin{eqnarray}\label{56}
\frac{dt}{C_{1t}(\rho,\Theta,\Phi)}&=&\frac{dr}{C_{1r}(\rho,\Theta,\Phi)}=
\frac{d\rho}{R_1(\rho,\Theta,\Phi)}\cr
\frac{dt}{C_{2t}(\rho,\Theta,\Phi)}&=&\frac{dr}{C_{2r}(\rho,\Theta,\Phi)}=
\frac{d\rho}{R_2(\rho,\Theta,\Phi)}\cr
\frac{dt}{C_{3t}(\rho,\Theta,\Phi,u^\mu)}&=&\frac{dr}{C_{3r}(\rho,\Theta,\Phi,u^\mu)}=
\frac{d\rho}{R_3(\rho,\Theta,\Phi,u^\mu)}.
\end{eqnarray}
give ``corpuscular-like'' motions along trajectories.

 Let me assume that one has two LDV's, say, Hamiltonian
$\vec{H}$ and momentum $\vec{P}$. Since $F^n_{1,2}(\pi, \Omega)$
consist in general all components of $\Omega^\alpha$, one should
take into account the commutation relations of the space-time
transformations accompanying the infinitesimal unitary
transformations when one tries to determine arbitrary functions
arising in PDE solutions.  We can build two field equations like
(\ref{PotEn}), (\ref{KinEn}). One of them for gauge field
representing homogeneous part of Lorentz group $E_1(\Omega)=0$ with
the solution $\Omega_{s1}=\Omega_{s1}(f_1(c^2t^2-r^2))$, containing
an arbitrary function $f_1(c^2t^2-r^2)$. The third one
$E_3(\Omega)=0$ has the solution
$\Omega_{s3}=\Omega_{s3}(f_3(r-t(u_x + u_y + u_z)))$, containing an
arbitrary function $f_3(r-t(u_x + u_y + u_z))$. Two nonequivalent
equations should have different solutions, but they concern the same
``field shell'' represented by the $\rho(t,r)$! Let me apply the
equation $E_3$ to the solution $\Omega_{s1}$. This yields
$E_3(\Omega_{s1}(f_1(c^2t^2-r^2)))=0$, then one will have a solution
for $f_{s1}=f_1(c^2t^2-r^2)$. Now let me apply the equation $E_1$ to
the solution $\Omega_{s3}$. This yields $E_1(\Omega_{s3}(f_3(r-t(u_x
+ u_y + u_z)))=0$, then one will have a solution for
$f_{3}=f_3(r-t(u_x + u_y + u_z))$, where \textbf{u} is rotated
\textbf{v}. One should find the condition of the self-consistency of
these equations generated by LDV's of Hamiltonian $\vec{H}$ and
momentum $\vec{P}$. I will discuss this problem elsewhere.

Field equations (\ref{KinEn}),(\ref{PotEn}) has relativistic
solutions $\Omega^\alpha$ for the coefficient functions for the
Hamiltonian flow (\ref{HVF}). Some solutions for the simplest case
$N=2$ have been found \cite{Le1,Le2}. The choice of the dimension $N
> 2$ gives different adjoint representation with dimension $N^2-1$.
The limit $N \to \infty$ is now interesting for our aim. This is a
difficult task to analyze the infinite set of so complicated
non-linear PDE system. But the increment of the dimensionality
sharply decreases amplitudes of all processes that effectively leads
to finite dimension problem. Nevertheless, some estimation is
encouraging.

The Fubini-Study metric gives us the possibility to measure distance
between adjacent rays corresponding original and deformed states in
$CP(N-1)$ or in $CP(\infty)$. In the last case one, however, should
be sure that the `transversal' part of the deformed vector belongs
to the tangent Hilbert space too. Then the natural measure is the
length $S$ of a curve. The derivative of a tangent vector
$\frac{dT^i(\pi)}{dS}$ generally is not tangent vector even in
finite dimension case, but the covariant derivative
$\frac{dT^i(\pi)}{dS} + \Gamma^i_{jk} T^k  \frac{d\pi^j}{dS}$ is the
tangent to $CP(N-1)$, where $\Gamma^i_{jk}$ is the affine
connection. One can treat this behavior of the tangent state vector
as a finite dimension non-linear model of the ``beat out'' of the
state vector due to the `divergences problem' \cite{Dirac2} from one
tangent Hilbert space to another (Dirac thought that it is ``beating
out'' to nowhere). Then we can assume that the real divergences
problem may be resolved in similar manner: in the infinite dimension
case there is smooth manifold $CP(\infty)$ and dynamical quantum
state (tangent vector) creeps from one tangent Hilbert space to
another. The convergence is required now only locally in the
specified tangent Hilbert space (in the sense of the Fubini-Study
metric). It should be noted that if $\sum_{n=0}^{\infty}|g^a|^2 <
\infty$, then each coherent state
$(\pi^1,...,\pi^k,...\pi^{N-1},...)$ belongs to the Hilbert space
$l^2$ since from the inequality, for example in the map $U_0 :\{|G>,
|g^0| \neq 0\}$, follows that $\sum_{n=0}^{\infty}|g^n|^2 = |g^0|^2
(1+\sum |\pi^s|^2)< \infty $, and, consequently,
$\sum_{i=1}^{\infty}|\pi^i|^2 < \infty $. Let us assume that tangent
vector field $T^i (\pi)$ belongs to $l^2$, i.e.
$\sum_{i=1}^{\infty}|T^i|^2 < \infty $. Let me put $\beta =
\sum_k^{\infty} |T^k|^2 \sum_s^{\infty} |\pi^s|^2 - \sum_k^{\infty}
|T^k \pi^k|^2 > 0$. Since for some $K < \infty $ one has $ 0 \leq L
=\sum_k^{\infty} |T^k|^2 \sum_s^{\infty} |\pi^s|^2 < K $ and $ 0
\leq M =\sum_k^{\infty} |T^k \pi^k|^2 < K $, and, $L - M
>0 $ one has $ \beta = L-M < L+M < 2K < \infty$, and, therefore,
$G_{ik*} T^i T^{k*} =\sum_k^{\infty} |T^k|^2(1+\sum_s^{\infty}
|\pi^s|^2)^{-2}(1+\frac{\beta}{\sum_k^{\infty} |T^k|^2}) < \infty $.

The requirement of the covariant derivative of the vector field $T^i
(\pi)$ elimination leads to the parallel transport equation $\frac{
\delta T^i (\pi)}{ \delta \tau}= \frac{ d T^i (\pi)}{ d \tau}+\Gamma
^i_{km} T^k \frac{ d\pi^m } { d \tau}=0$. Then for each fixed index
$i$ one has $\Lambda^i = \Gamma ^i_{km} T^k \frac{ d\pi^m }{ d \tau}
= (1/2) (T^i+\xi^i)\frac{\sum_{m=1}^\infty \pi^{m*}(T^m+\xi^m)}
{1+\sum_{s=1}^\infty |\pi^s|^2}$, where $\xi^m = \frac{d\pi^m }{ d
\tau}$. Because the operation of the vector summation in $l^2$ is
correct, i.e. $\sum_{s=1}^\infty |T^s + \xi^s|^2 < \infty $, if $T,
\quad \xi \in l^2$, and because $\alpha_m = \frac{|\pi^m|}
{1+\sum_{s=1}^\infty |\pi^s|^2} < 1$, one has the inequality
$\sum_{m=1}^\infty \alpha_m^2 |T^m+\xi^m|^2 < \infty $.
Nevertheless, one has not guaranty that any sequence of $\Lambda^i $
belongs to the $l^2$. But there is a possibility to find such
$\delta_1, \Delta_1$ that for $ \sum_{s=1}^\infty |\pi^s|^2 <
\delta_1$ or for $ \sum_{s=1}^\infty |\pi^s|^2 > \Delta_1 $ we will
get $\sum_{s=1}^\infty |\Lambda^s|^2 < \infty $.

\vskip 0.2cm I am grateful to L.P.Horwitz for a lot of interesting
discussions and critical notes.

\end{document}